\documentclass[printer]{aa}

\usepackage{txfonts}
\usepackage{graphicx}
\usepackage{lscape}
\usepackage{natbib}
\bibpunct{(}{)}{;}{a}{}{,} 

\newcommand{\sourcename}{J1409+5628}
\newcommand{\kms}{km s$^{-1}$}

\newcommand{\Kkmspc}{K km s$^{-1}$  pc$^2$}

\newcommand{\Lsun}{L$_\odot$}
\newcommand{\Msun}{M$_\odot$}
\newcommand{\LprimeCO}[2]{L^\prime_{\rm CO(#1 \rightarrow #2)}}

\def\to {$\rightarrow$}

\begin{document}

\title{Starburst activity in the host galaxy of the $z=2.58$ quasar J1409+5628
\thanks{This paper is based on observations obtained with the IRAM Plateau
  de Bure Interferometer.  IRAM is funded by Centre National de la
  Recherche Scientifique (France), the Max-Planck-Gesellschaft
  (Germany), and the Instituto Geografico Nacional (Spain).}}

\author{A. Beelen\inst{1}, 
        P. Cox\inst{1}, 
        J. Pety\inst{2}, 
        C.L. Carilli\inst{3}, 
        F. Bertoldi\inst{4}, 
        E. Momjian\inst{5}, 
        A. Omont\inst{6}, 
        P. Petitjean\inst{6}, 
        A.O. Petric\inst{7} 
}

\institute{Institut d'Astrophysique Spatiale, 
           Universit\'e de Paris XI, F-91405 Orsay, France
      \and IRAM,  300 rue de la Piscine, 38406 St-Martin-d'H\`eres, France  
      \and National Radio Astronomy Observatory, P.O. Box,
           Socorro, NM~87801, USA
       \and Max-Planck-Institut f\"ur Radioastronomie, Auf dem H\"ugel 69,
           D-53121 Bonn, Germany     
       \and  NAIC, Arecibo Observatory, HC 3 Box 53995, Arecibo, PR 00612, USA
       \and Institut d'Astrophysique de Paris, CNRS \& 
           Universit\'e Paris 6, 98bis bd. Arago, 75014 Paris, France     
       \and Astronomy Department, Columbia University, New York, NY USA    
     }

\offprints{A. Beelen, \email{Alexandre.Beelen@ias.u-psud.fr}}

\date{Received 23 February 2004 / Accepted **** }

\titlerunning{CO and radio emission in a quasar at redshift 2.58}
\authorrunning{A. Beelen et al.}

\abstract{We report the detection of CO emission from the optically
  luminous, radio-quiet quasar J140955.5+562827 (hereafter
  \sourcename), at a redshift $z_{\rm CO} =2.583$.  We also present
  VLA continuum maps and VLBA high spatial resolution observations at
  1.4~GHz. Both the CO(3\to2) and CO(7\to6) emission lines are
  detected using the IRAM Plateau de Bure interferometer. The
  3\to2/7\to6 line luminosity ratio is about 1/3, indicating the
  presence of warm and dense molecular gas with an estimated mass of
  $6 \times 10^{10}$~\Msun.  The infrared-to-CO luminosity ratio
  $L_{\rm FIR}/L^\prime_{\rm CO(1\rightarrow0)} \approx
  500$~\Lsun~(K~km~s$^{-1}$~pc$^2$)$^{-1}$, comparable to values found
  for other high-$z$ sources where CO line emission is seen.
  \sourcename~ is detected using the VLA with a 1.4~GHz rest-frame
  luminosity density of $4.0\times10^{25} \, \rm W \, Hz^{-1}$.  The
  rest-frame radio to far-infrared ratio, $q$, has a value of 2.0
  which is similar to the values found in star forming galaxies.  At
  the 30~mas resolution of the VLBA, \sourcename~ is not detected with
  a 4$\sigma$ upper limit to the surface brightness of
  0.29~mJy~beam$^{-1}$. This implies a limit to the intrinsic
  brightness temperature of $\rm 2 \times 10^5 \, K$ at 8~GHz, typical
  for nuclear starbursts and more than two orders of magnitude fainter
  than typical radio-loud active galactic nuclei. Both the properties
  of the CO line emission and the radio emission from \sourcename~ are
  therefore consistent with those expected for a star forming galaxy.
  In \sourcename~ young massive stars are the dominant source of dust
  heating, accounting for most of the infrared luminosity. The massive
  reservoir of molecular gas can sustain the star formation rate of a
  few $\rm 1000 \, M_\odot \, yr^{-1}$ implied by the far-infrared
  luminosity for about 10 million years.

}

\maketitle \sloppy

\section{Introduction}
\label{sec:Introduction}

In the recent decade, millimeter and submillimeter deep blank field
surveys \citep[e.g.,][]{Ivison2000,Blain2002} and pointed observations
\citep[e.g.,][]{Omont2003,Priddey2003,Bertoldi2003} have provided a
view of the dust content of a few hundred galaxies and quasars in the
redshift range $1 < z < 6.4$.  In these objects, a significant
fraction of the energy generated by star formation is processed by
dust and re-emitted at far-infrared wavelengths. The average space
density of ultraluminous infrared galaxies and quasars at high-$z$ is
found to be thousandfold greater than in the local universe.
Observations at millimeter and submillimeter wavelengths provide a
direct way to trace the bulk of the star formation in the early
universe.
  
Pointed submillimeter and millimeter observations of optically
selected quasars or radio galaxies have the advantage over deep field
surveys in providing unique source identification and redshifts.  In
particular, targeted observations of optically luminous quasars have
been successful in revealing massive bursts of star formation in their
host galaxies. Although optically luminous quasars, which trace the
most massive collapsed structures to have formed in the early
universe, are rare objects which bear little relation to the objects
that make up most of the far-infrared and submillimeter background,
they provide powerful probes to study the relation between star
formation, massive gas reservoirs, and the growth of super-massive
black holes in the dark ages of the Universe
\citep{Fan2003,Bertoldi2003b,Walter2003}.

The study of the redshift range $2 < z <3$ is important in studying
the star formation history in the universe since it traces the peak of
the space density of the quasar population \citep{Shaver1996}, and
corresponds to the median redshift of submillimeter galaxies
\citep{Chapman2003}.  Recent pointed observations of the 1.2 and
0.85~mm thermal dust continuum emission of $z\approx2$ optically
luminous, radio-quiet quasars reveal that 1/3 of these quasars are
also luminous in the infrared with far-infrared luminosities of
$L_{\rm FIR} \rm \sim 10^{13}$~\Lsun~ and estimated dust masses of
typically $\rm \sim 10^8 \, M_\odot$ \citep{Omont2003,Priddey2003}.

The optically very bright ($M_B=-28.4$), radio-quiet quasar
\sourcename~ is by far the strongest mm source in the
\citet{Omont2003} survey. With an estimated far-infrared luminosity of
$4 \times 10^{13}$~\Lsun, \sourcename~ ranges amongst the most
luminous infrared high-$z$ sources found to date.  Under the
assumption that the dust is predominantly heated by massive stars, the
inferred star formation rate is several $\rm 1000 \, M_\odot \,
yr^{-1}$ \citep{Omont2003}.  However, the fraction of infrared
luminosity due to dust heated by star formation and dust directly
heated by the Active Galactic Nuclei (AGN) in \sourcename~ is still an
open issue, as is the case for almost all of the high-$z$ quasars
studied to date.  Both the large masses of dust and the high star
formation rate indicate that \sourcename~ should have a copious
reservoir of molecular gas to sustain the formation of a few $1000 \,
M_\odot$ per year over its dynamical time.

In high-$z$ sources, the search for molecular gas is best done by
using the redshifted rotational lines of CO. The molecular gas in
star-forming galaxies is warm and dense enough to excite the higher CO
rotational levels ($\rm J>3$) which are shifted into the millimeter
windows for redshifts greater than 2.  However, a good determination
of the redshift of the target object, either from optical or
near-infrared spectroscopy, is a prerequisite for a detection because
the bandwidths of the current heterodyne receivers are still
relatively narrow. For example, at the Plateau de Bure Interferometer,
the available instantaneous bandwitdh is $\Delta \nu = 580$~MHz which
corresponds at 96~GHz to a velocity range of 1810~\kms or $\Delta z
\sim 0.02$ at $z=2.6$.

Despite this current limitation, over the last few years CO line
emission has been detected in 24 far-infrared luminous high-$z$
quasars, radio and (sub)mm galaxies \citep[][ and references
therein]{Carilli2004}.  These observations reveal large reservoirs of
molecular gas ($\rm 10^{11} \, M_\odot$), a prerequisite for efficient
star formation, and prompted also the first studies on the properties
of the dense interstellar gas in starburst galaxies in the early
universe \citep[e.g.][]{Barvainis1997,Solomon2003,Bertoldi2003b}.  In
addition, these observations allow us to estimate, unhindered by
extinction, the dynamical masses in these systems, which are the key
constraining hierarchical models \citep[see, e.g.][]{Neri2003}.

In this paper, we present a study of the CO emission in \sourcename~
made with the Plateau de Bure interferometer together with
observations of the radio emission carried out with the Very Large
Array and the Very Large Baseline Array. The data provide strong
evidence that the origin of the infrared luminosity is dominated by
the starburst activity of the host galaxy of the QSO \sourcename.
Throughout this paper, we adopt the concordance $\Lambda$-cosmology
with $H_0=71\rm~km~s^{-1}~Mpc^{-1}$, $\Omega_\Lambda=0.73$ and
$\Omega_m=0.27$ -- \citep{Spergel2003}. At $z=2.58$, the luminosity
distance $D_L$ is 21.5~Gpc and an angular scale of $1^{\prime\prime}$
corresponds to 8.1~kpc.

\section{Observations}
\label{sec:observations}

\subsection{Plateau de Bure Interferometer}
\label{sec:PdBI}
 
Observations were made with the IRAM Plateau de Bure interferometer in
May and June 2002 and from January to March 2003. We used the D and B
configurations, as well as non-standard configurations, with 4 to 6
antennas, resulting in a good coverage of the \emph{uv} plane with
baselines ranging from 24 to 400 meters. This results in final
synthesized beams of $2\farcs44 \times 1\farcs65$ at a position angle
of $60\degr$ at 3.1~mm, and $1.01\arcsec \times 0.55~\arcsec$ with a
position angle of $52\degr$ at 1.3~mm.

The dual frequency set-up was used to search simultaneously for the
CO(3\to2) emission line and dust emission at 3 and 1.3~mm, and
subsequently for the CO(7\to6) emission line.  During the first series
of observations, the 3.1~mm receiver was tuned to a central frequency
of 96.780~GHz, corresponding to the CO(3\to2) line at a redshift of
$z=2.5730$, close to the strong Lyman-$\alpha$ absorption line seen at
$z=2.5758$ in the optical spectrum of \sourcename~(see
Sect.~\ref{sec:redshift}).  The CO(3\to2) line emission was
tentatively detected at the low frequency end of the available
bandwidth.  In subsequent observations, we re-tuned the 3.1~mm
receivers to 96.505~GHz, corresponding to a redshift of $z=2.583$ and
confirmed this tentative detection.  After the first tentative
detection of the CO(3\to2) emission line, we retuned the 1.3~mm
receivers in the upper sideband to 225.198~GHz corresponding to the
redshifted CO(7\to6) emission line, while the lower sideband was tuned
to a frequency 3~GHz away (at 225~GHz) to measure the 1.3 mm
continuum.  Typical SSB system temperatures were $\approx 150$~K and
$\approx 400$~K at 3~mm and 1.3~mm respectively.

The data were reduced, calibrated and analyzed using the IRAM package
{\sc Gildas}. During the data reduction, we flagged antennas with too
low efficiencies or too high system temperatures, and discarded data
taken during bad atmospheric conditions.  The final usable datasets
correspond to total on-source integration times of 52 and 27~h at 3.1
and 1.3~mm, respectively.  After phase and amplitude calibration of
each session, we combined the data from the different settings into
one single table at full spectral resolution. The flux calibration
was done using standard calibrators 3C~273, CRL~ 618 or MWC~349. 

\begin{figure*}[htb]
  \resizebox{\hsize}{!}{\includegraphics[angle=-90,width=1\linewidth]{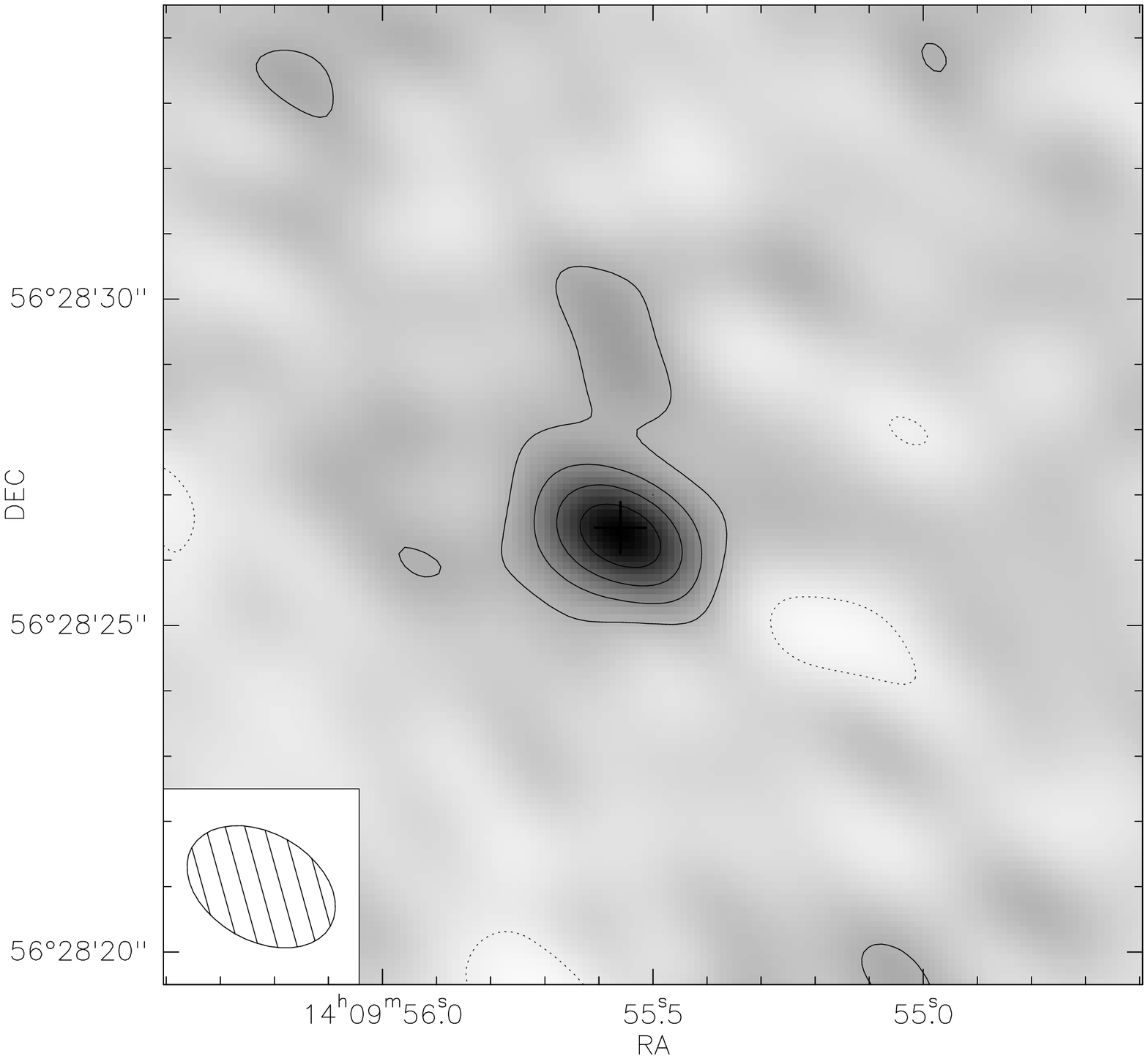},
    \includegraphics[angle=-90,width=1\linewidth]{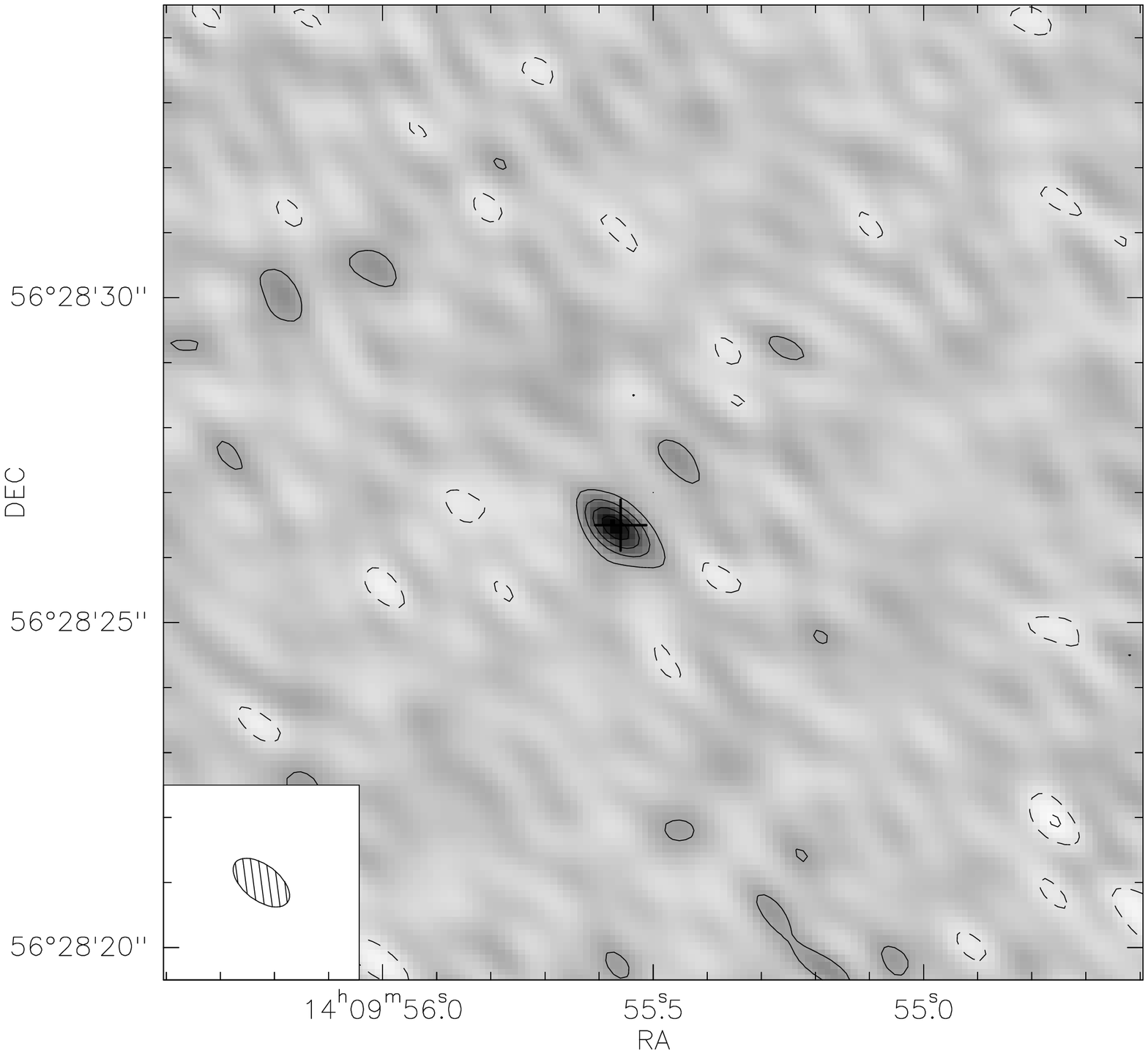}}
\caption{\emph{Left panel}: Velocity-integrated map of the 
  CO(3\to2) line emission toward J~1409+5628.  The contour step is
  0.3~mJy/beam, corresponding to $2\sigma$, dashed contours are
  negative.  \emph{Right panel} : Velocity-integrated map of the
  CO(7\to6) line and 1.3~mm continuum emission toward J~1409+5628.
  The contour step is 2~mJy/beam, corresponding to $2\sigma$, dashed
  contours are negative. In both panels, the cross shows the optical
  position of the quasar taken from the 2MASS catalogue. The
  synthesized beams are shown in the lower left corner of each panel.}
\label{fig:map} 
\end{figure*} 

\subsection{Very Large Array}
\label{sec:VLA}

\sourcename~ was observed with the Very Large Array (VLA) of the
NRAO\footnote{ The National Radio Astronomy Observatory is a facility
  of the National Science Foundation operated under cooperative
  agreement by Associated Universities, Inc.} at 1.4~GHz in the A
configuration (maximum baseline of 30 km), with a total bandwidth of
100~MHz and two polarizations. The total observing time was 2 hours.
The entire dataset was self-calibrated using field sources, and
standard phase and amplitude calibration was applied for \sourcename.
The absolute flux density scale was set with the observations of
3C~286.

The final image was generated using the wide field imaging and
deconvolution capabilities of the NRAO's Astronomical Image Processing
System (AIPS) task \emph{IMAGR}. The present 1.4~GHz continuum data
achieved a rms noise ($\sigma$) of $\rm 16 \, \mu Jy$ which
corresponds to the theoretical expected noise level. The Gaussian
restoring CLEAN beam full width at half-maximum (FWHM) is $\sim
1\farcs5$. A detailed discussion of the VLA data will be given in
Petric et al. (in preparation).

\subsection{Very Long Baseline Array}
\label{sec:VLBA}

We observed \sourcename~ with the VLBA of the NRAO on the 9th of March
and 7th of April, 2003, for a total of 14 hours using the standard
continuum mode at 1.4~GHz (total bandwidth of 16~MHz in two dual
circular polarizations). The calibrator J1408+5613 was used for phase
referencing with a calibration cycle time of 3.5 minutes. Standard
\textsl{a priori} gain calibration was applied. At 1.4~GHz, the VLBA
short spacing limit filters out all spatial structure larger than
about $0\farcs15$.

A number of test cycles were also included to monitor the coherence of
the phase referencing.  These tests involved switching between two
calibrators (J1408+5613 and J1419+5423) with a similar angular
separation and cycle time as that used for the target source.  Images
of the second calibrator (J1419+5423) were deconvolved using two
different approaches: by applying the phase and amplitude
self-calibration solutions of the phase reference source J1408+5613 on
J1419+5423, and by self calibrating J1419+5423 itself, both in phase
and amplitude. The ratio of the peak surface brightness between the
final images of the two approaches gives a measure of the effect of
residual phase errors after phase referencing (i.e. 'the coherence'
due to phase referencing).  At all times the coherence was found to be
better than 95$\%$. Also, phase referencing as used herein is known to
preserve absolute astrometric positions to better than $0\farcs1$
\citep{Fomalont1999}.

\section{Results}
\label{sec:results}

\subsection{Molecular gas} 
\label{sec:results:molecular_gas}

\begin{table*}[htp]

\caption{Properties of the CO lines observed toward J1409+5628}
\begin{center}
\begin{tabular}{ccccccccc}
\hline
\hline 
Line                 & $\nu_{\rm obs}$ & $z_{\rm CO}$       & peak int. & $\Delta \rm v_{\rm FWHM}$ & $S_{\nu}$          & $I_{\rm CO}$     & $L^\prime_{\rm CO}$              & $L_{\rm CO}$        \\
                     &       [GHz]     &                    & [mJy]     & [km s$^{-1}$]         &  [mJy]             & [Jy km s$^{-1}$] & [$10^{10}$~K km s$^{-1}$ pc$^2$] & [$10^8$~ L$_\odot$] \\ 

\hline \\[-0.2cm]

CO(3\to2)           & $96.504$         & $2.5832\pm0.0001$  & $6\pm1$  &  $311\pm28$            &  $<0.5^\dagger$    & $2.3\pm0.2$      &  $8.2\pm0.6$                     & $1.1\pm0.1$         \\ 

CO(7\to6)           & $225.1197$       & $^\diamond$        & $10\pm3$  &  $^\diamond$          &  $6\pm2$           & $4.1\pm1.0$      &  $2.6\pm0.7$                     & $4.4\pm1.2$         \\

\hline
\end{tabular}
\end{center}
\label{tab:properties}

Notes --  
$^\dagger$ 3$\sigma$ level upper limit.
$^\diamond$ Adapting the CO(3\to2) parameters (see text)

\end{table*}


\begin{figure*}[tbp!]
\sidecaption
 \includegraphics[width=1\linewidth]{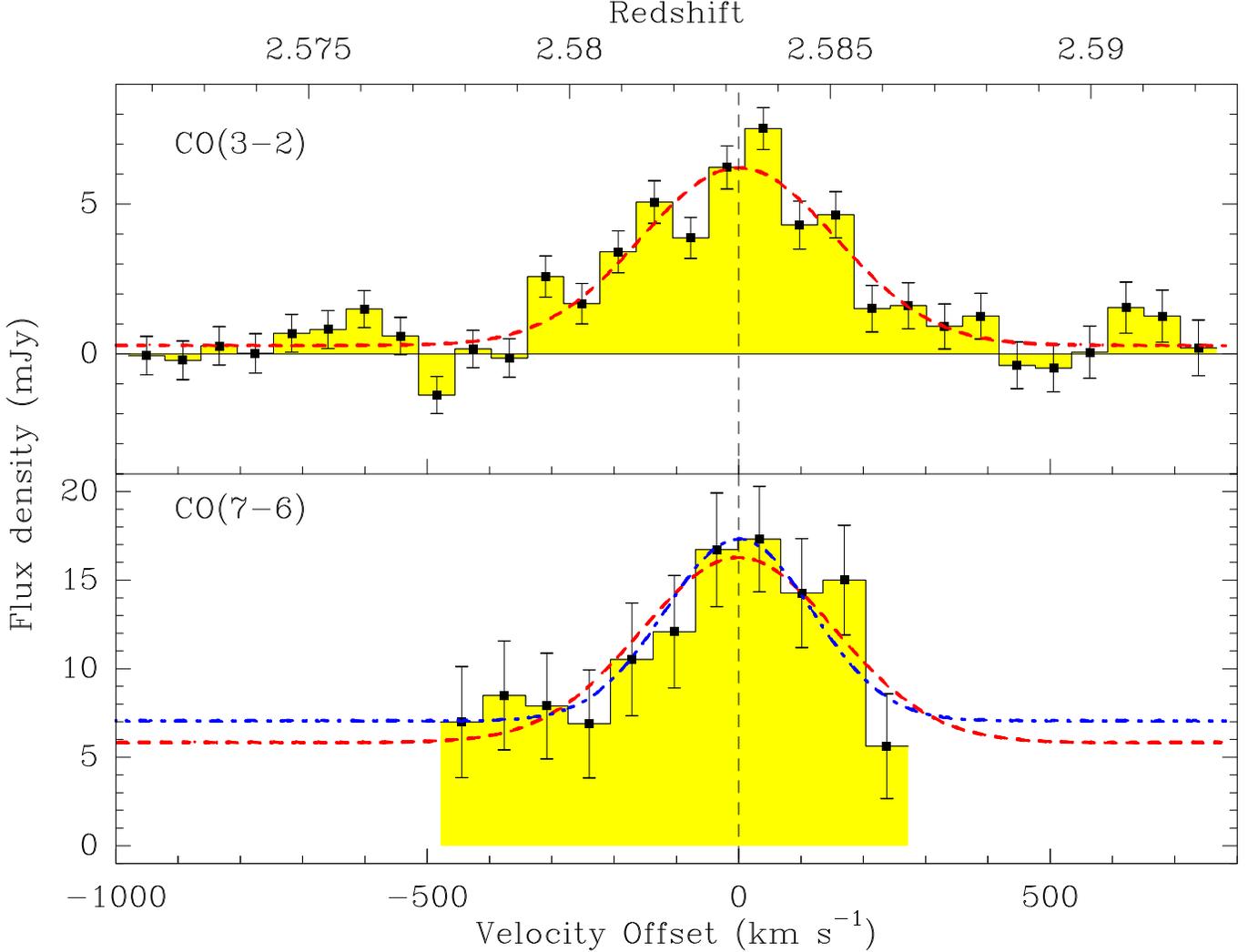}
    \caption{Observed spectra of the CO(3\to2) and CO(7\to6) line emission 
      toward \sourcename, with velocity resolutions of 58 and 76~\kms,
      respectively. The offset velocity corresponds to central
      frequencies of 96.504 and 225.1197~GHz.  The error bars
      represent the statistical uncertainties from the fit in the
      \emph{uv}-plane.  The dashed lines show the gaussian fits
      together with the fit to the continuum with the parameters
      listed in Table~\ref{tab:properties}. The dot-dashed line is the
      fit to the CO(7\to6) line emission leaving the linewidth as a
      free parameter }
  \label{fig:spectrum}
\end{figure*} 

The velocity-integrated maps of the CO(3\to2) and CO(7\to6) emission
lines of \sourcename~ are shown in Fig.~\ref{fig:map} and the
corresponding spectra are displayed in Fig.~\ref{fig:spectrum}.  The
CO(3\to2) line is detected with high signal-to-noise ratio at a
central frequency of $96.504$~GHz, corresponding to a redshift $z_{\rm
  CO} = 2.5832\pm0.0001$.  At the same redshift, the CO(7\to6)
emission line is also detected on top of the 1.3~mm continuum.

Within the astrometric uncertainties, the CO emission and the 1.3~mm
continuum, centered at $\rm 14^h 09^m 55\fs5 \, +56\degr 28\arcmin
26\farcs4$ (J2000.0), coincide with the optical position listed in the
2MASS catalogue \citep{Barkhouse2001}.  At the $1.0\arcsec \times
0.5\arcsec$ resolution of our Plateau de Bure observations at 1.3~mm,
the CO(7\to6) emission in \sourcename~ is still unresolved, yielding
an upper limit to the linear size of the region emitting in CO of
5~kpc.

We fitted the high signal-to-noise CO(3\to2) spectrum with a gaussian
profile where the central frequency ($\nu_{\rm obs}$), the full width
half maximum ($\Delta \rm v_{\rm FWHM}$), the continuum flux density
($S_{\nu}$), and the integrated CO emission flux density ($I_{\rm
  CO}$) were kept as free parameters. The peak intensity was derived
from the gaussian fit parameters.  The best gaussian fit yields a line
width to the CO(3\to2) emission of $\rm 311 \pm 28$~\kms, similar to
the widths found in other high-$z$ quasars \citep[e.g.,][]{Cox2002},
and a peak intensity of $6 \pm 1$~mJy. The velocity-integrated
CO(3\to2) flux is $2.3 \pm 0.2 \, \rm Jy \, km \, s^{-1}$.  At the
position of \sourcename~ we obtain from the combined fit a $3\sigma$
upper limit to the continuum flux density at 3.1~mm of $S_{\rm 96.5
  GHz} < 0.5$~mJy.

The width and centroid of the gaussian fit to the CO(3\to2) line were
adopted for the profile of the lower signal-to-noise ratio CO(7\to6)
emission line.  The peak intensity of the CO(7\to6) emission is
$10\pm3$~mJy, the velocity-integrated CO(7\to6) flux $4.1 \pm 1.0 \,
\rm Jy \, km \, s^{-1}$, and the 1.3~mm continuum is detected at a
flux density of $\rm 6\pm2 \, mJy$.  The CO(7\to6) emission line is
detected with a signal-to-noise ratio of 4, as compared to the ratio
of 11 in the case of CO(3\to2). The CO(7\to6) line flux is therefore
somewhat uncertain, and the precise value depends on how the spectrum
is analyzed. For instance, leaving the line width as a free parameter
in the gaussian fitting yields $\Delta \rm v_{\rm FWHM} = 229 \pm 84
\, \rm km \, s^{-1}$ and a CO(7\to6) integrated line flux of $3.0 \pm
1.3 \, \rm Jy \, km \, s^{-1}$, i.e. a 2.3~$\sigma$ result
(Fig.~\ref{fig:spectrum}).

The upper limit of the continuum flux density at 96~GHz and the
continuum flux density at 225~GHz are consistent with the 250~GHz flux
density of $S_{\rm 250\ GHz} = 10.7 \pm 0.6$~mJy \citep{Omont2003},
for a grey-body spectrum with a dust temperature of 45~K and a
spectral index $\beta=1.5$. The corresponding far-infrared luminosity
is $L_{\rm FIR} \approx 4.3 \times 10^{13}$~\Lsun. The recent
detection of \sourcename~ at 350~$\mu$m is in agreement with these
conclusions (Beelen et al. in preparation).

\subsection{Non-thermal radio emission} 
\label{sec:results:radio}

\begin{figure}[htb!]
   \resizebox{\hsize}{!}{\includegraphics[angle=-90]{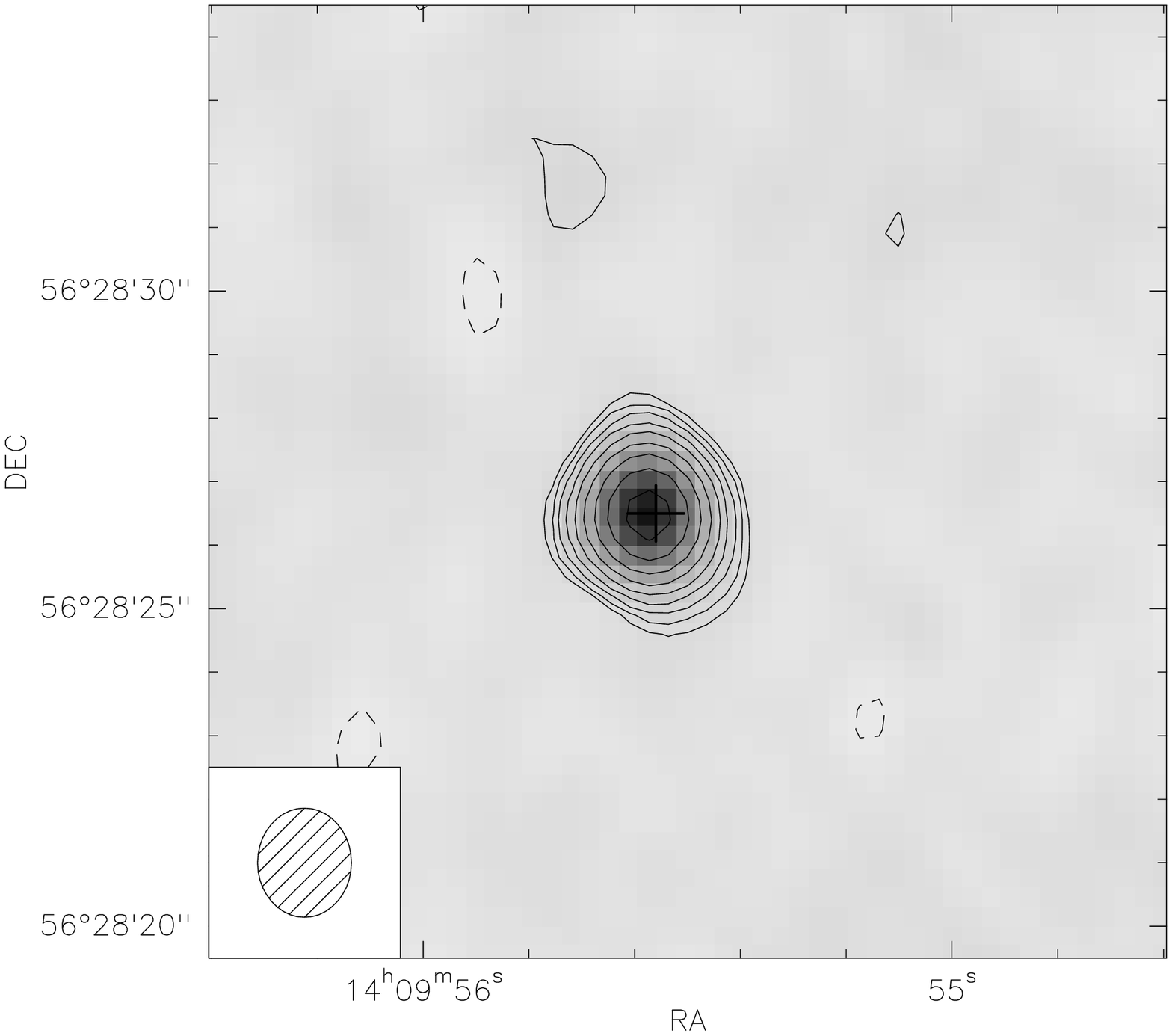}}
   \caption{VLA image of \sourcename~ at 1.4~GHz at a resolution
     of $1\farcs7 \times 1\farcs5$ (FWHM), major axis position angle
     of 5$\degr$.  The contours are a geometric progression in the
     square root of two, with the first contour level at 0.05
     mJy/beam, and the highest one at 128~$\sigma$.  Negative contours
     are dashed.}
\label{fig:radio_vla} 
\end{figure}

\begin{figure*}[htb!]
   \resizebox{\hsize}{!}{\includegraphics[height=1\linewidth]{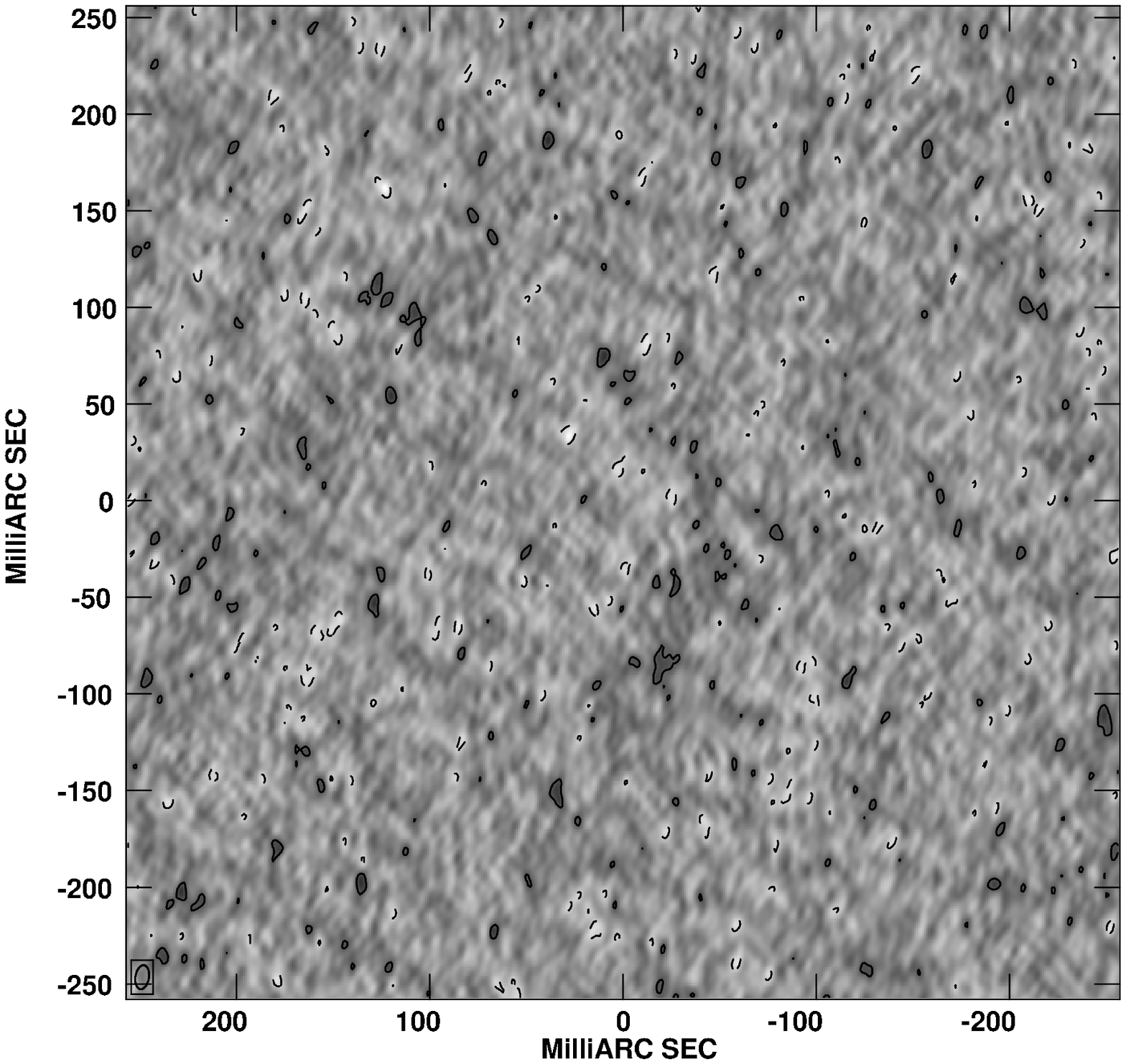},
     \includegraphics[height=1\linewidth]{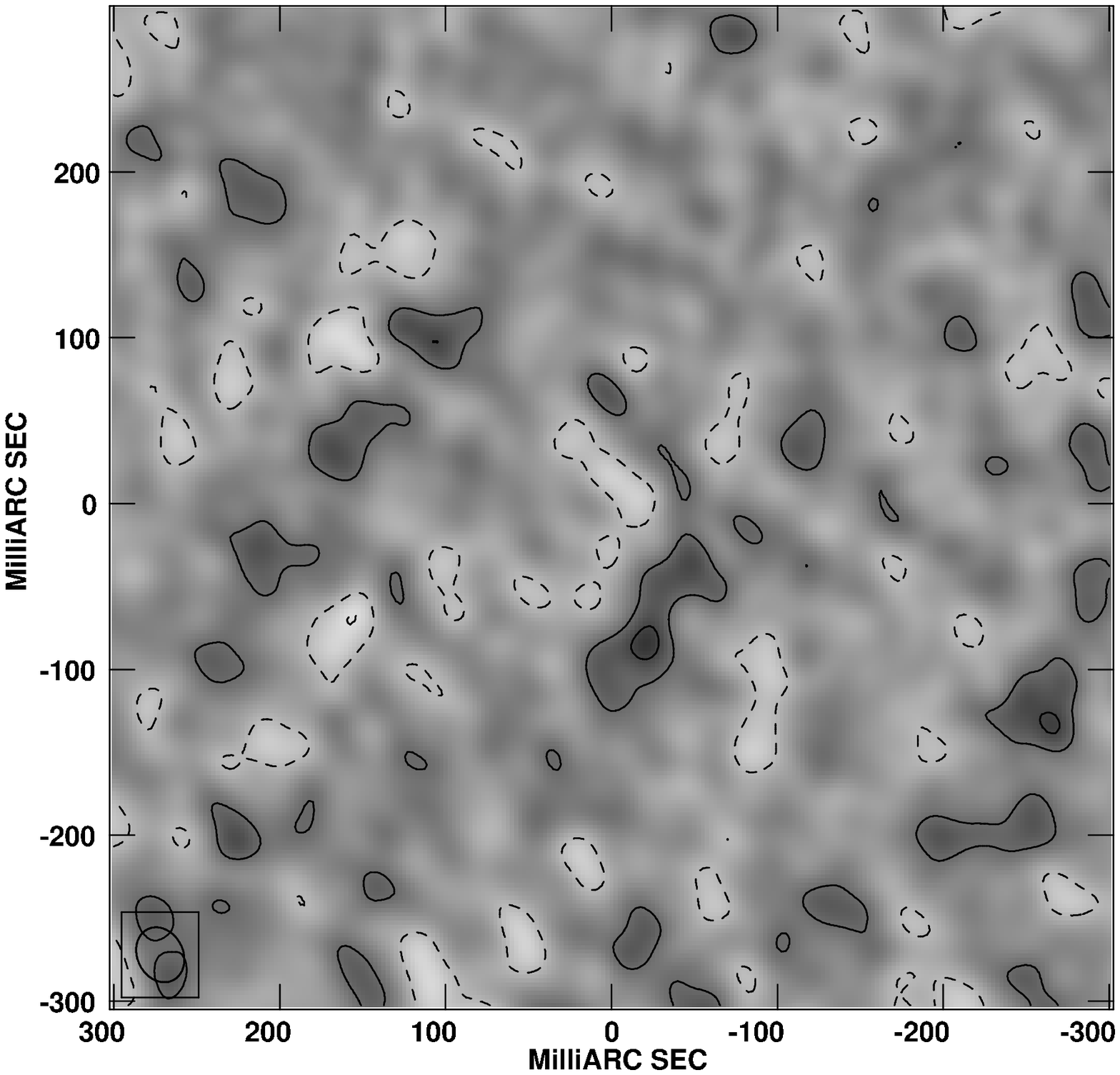}}
   \caption{\emph{Left panel} VLBA 1.4~GHz image at full
     resolution (12.5mas $\times$ 6.7 mas, major axis position angle
     $-9\degr$). The contours levels are: -0.3, -0.2, -0.1, 0.1, 0.2,
     0.3, 0.4 mJy beam$^{-1}$.  \emph{Right panel} The VLBA 1.4~GHz
     image tapered to $34 \times 28$ mas resolution.  The contours
     are: $-0.33$, $-0.22$, $-0.11$, 0.11, 0.22, 0.33,
     0.44~mJy~beam$^{-1}$. The central position corresponds to $\rm
     14^h 09^m 55\fs5739$, $\rm +56\degr 28\arcmin 26\farcs475$
     (J2000).}
\label{fig:radio_vlba} 
\end{figure*}

Figures~\ref{fig:radio_vla} and \ref{fig:radio_vlba} show the VLA and
VLBA radio images of \sourcename~ at 1.4 GHz.  The VLA image at
$1\farcs6$ resolution shows an unresolved source with a deconvolved
size $< 0\farcs5$, consistent with the upper limit derived from the CO
measurements. The flux density at 1.4~GHz is $\rm S_{\rm 1.4 \, GHz} =
0.93\pm 0.022$~mJy.  The radio source position is $\rm 14^h 09^m
55\fs57$ $+56\degr 28\arcmin 26\farcs47$ (J2000), and coincides within
the astrometric errors ($\le 0\farcs2$) with the positions of the
optical and the CO emission.

The VLBA image at full resolution ($\rm 12.5\,mas \times 6.7 \, mas$,
major axis position angle $-9\degr$; left panel of
Fig.~\ref{fig:radio_vlba}) has an rms noise of $46 \, \mu$Jy
beam$^{-1}$.  There is no source brighter than 4$\sigma$ in a field of
$0\farcs5$ centered on the VLA position.  We have also made tapered
images at lower resolution and wider fields.  At $34\times 28$~mas
resolution the rms noise is $\rm 72 \, \mu Jy \, beam^{-1}$. The
tapered image is shown in the right panel of
Fig.~\ref{fig:radio_vlba}.  We still find no source brighter than
4$\sigma$ within $0\farcs3$ of the field center. Near the center of
the field there is a 'linear' feature extending about $0\farcs1$
north-south with a peak surface brightness of $0.27\pm 0.07$ mJy
beam$^{-1}$ at $\rm 14^h 09^m 55\fs572$, $+56\degr 28\arcmin
26\farcs39$ (J2000).  It is possible that this linear feature
corresponds to the \sourcename~ radio source, but given the low
surface brightness, we conclude that our VLBA observations can only
set a 4$\sigma$ upper limit to the surface brightness at 30~mas
resolution of 0.29~mJy beam$^{-1}$.

\section{Discussion}
\label{sec:properties}

\subsection{The redshift of \sourcename}
\label{sec:redshift}

The redshift derived from the CO measurements should correspond to the
systemic redshift of the host galaxy of \sourcename, since it traces
the extended molecular gas associated with the quasar and not ionized
gas undergoing energetic processes linked with the AGN activity.  The
derived redshift for the molecular gas, $z_{\rm CO} = 2.5832 \pm
0.0001$, is in the upper part of the redshift range determined from
the optical spectra (\citealt{Korista1993}; \citealt{Barlow1994}),
i.e. $2.550 < z < 2.58$.  $z_{\rm CO}$ is close to the redshift of the
strong Ly-$\alpha$ absorption ($z = 2.5758$) seen in the Keck spectrum
\citep{Korista1993}.  This absorption line is associated with narrow
C~{\sc iv} absorption at 4352~\AA, indicating that the corresponding
gas is not part of the Lyman-$\alpha$ forest, but rather part of the
host-galaxy. This is in agreement with the fact that this ionization
state is not associated with the central AGN (see, e.g.,
\citealt{Petitjean1994}) and that its redshift is close to $z_{\rm
  CO}$.  The (systemic) velocity corresponding to $z_{\rm CO}$ is
redshifted by 1800~\kms from the $z=2.5624$ high ionization
ultraviolet emission lines such as C~{\sc iv} (\citealt{Korista1993}),
a difference which has been found in many other quasars
\citep{Richards2002}.

\subsection{Molecular gas emission}
\label{sec:properties:molecular_gas}

The $L^{\prime}_{\rm CO(7 \rightarrow 6)}$ luminosity is about 1/3 of
the $L^{\prime}_{\rm CO(3 \rightarrow 2)}$ luminosity
(Table~\ref{tab:properties}), a ratio which is comparable within a
factor of 2 to the 7\to6/3\to2 luminosity ratios measured in the
high-$z$ sources SMM~J14011+0252 \citep[0.2,][]{2003ApJ...582...37D},
H1413+117, i.e. the Cloverleaf \citep[0.8,][]{Barvainis1997}, and
J1148+5251 \citep[0.6,][]{Bertoldi2003b}.  As in the other sources,
the detection of CO(7\to6) emission in \sourcename~ indicates the
presence of warm and dense molecular gas.

Assuming a constant brightness temperature from CO(3\to2) to CO(1\to0)
and a conversion factor from $L^\prime_{\rm CO\ (1\rightarrow 0)}$ to
molecular mass of $X_{\rm CO} \approx 0.8$~\Msun~(\Kkmspc)$^{-1}$,
appropriate for ultraluminous galaxies \citep{Downes1998}, the CO
(3\to2) luminosity implies a mass of molecular gas of $M({\rm H_2})
\approx 6.6 \times 10^{10}$~\Msun.  \sourcename~ appears therefore to
be a system which is rich in molecular gas and comparable to the other
high-$z$ quasars detected in CO.

\subsection{Radio continuum emission}
\label{sec:discussion:radio}

Assuming a low frequency spectral index of $-0.8$, the 1.4~GHz
continuum flux density implies a rest frame luminosity density at
1.4~GHz of $4.0\times10^{25} \, \rm W \, Hz^{-1}$.  Using the
far-infrared luminosity definition of \citet{Condon1992}, extrapolated
from the 250~GHz flux density of \citet{Omont2003}, a dust temperature
of 45~ K and a spectral index $\beta=1.5$, the radio-to-FIR ratio is
then $q = 2.01$, where $q$ is defined by \citet{Condon1992}.  A tight
and linear correlation between radio and far-infrared luminosity has
been found in star forming galaxies, with small scatter over a few
orders of magnitude in luminosity \citep{Condon1992}.  The most recent
consideration of this correlation found a value of $q = 2.3 \pm 0.3$
for star forming galaxies from the IRAS 2~Jy sample \citep{Yun2001}.
Hence, the value of $q$ for \sourcename~ falls well within the
envelope defined by star forming galaxies.

\citet{Condon1991} derived an empirical upper limit to the brightness
temperature for nuclear starbursts of order $10^5$~K at 8~GHz, while
typical radio loud AGN have brightness temperatures two or more orders
of magnitude larger than this value.  They also present a possible
physical model for this limit involving a mixed non-thermal and
thermal radio emitting (and absorbing) plasma, constrained by the
radio-to-FIR correlation for star forming galaxies. For \sourcename,
we use the observed surface brightness limit from the VLBA together
with the measured spectral index and redshift to derive a 4$\sigma$
upper limit to the intrinsic brightness temperature of $2\times10^5$~K
at 8~GHz typical of starburst galaxies.  This is in contrast with the
results obtained by \citet{Momjian2004} for a sample of three high-$z$
quasars which were imaged with the VLBA. In these $z>4$ quasars, the
radio-loud AGN dominates the radio emission on very compact size, i.e.
a few milliarcsec, with intrinsic brightness temperatures in excess of
$10^9$~K.

\subsection{Starburst activity}
\label{sec:starburst}

Overall, the properties of the CO emission and the radio emission from
\sourcename~ are consistent with those expected for a star forming
galaxy.  The radio emission indicates a $q$ value which is consistent
with the radio-to-FIR correlation of star forming galaxies, and an
intrinsic brightness temperature below $2\times10^5$~K at 8~GHz
typical of nuclear starbursts. 

The ratio between the far-infrared and CO(1\to0) luminosities of
\sourcename~ implied by the present observations, $L_{\rm
  FIR}/L^\prime_{\rm CO(1\rightarrow0)} \approx
500$~\Lsun~(K~km~s$^{-1}$~pc$^2$)$^{-1}$ or $L_{\rm FIR}/L_{\rm
  CO(1\rightarrow0)} \approx 1.1\times 10^7$ is at the upper end of
the values derived for local ultraluminous infrared galaxies by
\citet{Solomon1997}, and in agreement with other high-$z$ quasars
where CO line emission is detected \citep{Cox2002}.

Following \citet{Omont2001}, the star formation rate (SFR) in
\sourcename~ can be derived from its far-infrared luminosity, assuming
that star formation is the main contribution to the dust heating.
Depending on the stellar initial mass function and the starburst age
and duration, the ratio of SFR to infrared luminosity is in the range
0.8 to $\rm 2 \times 10^{-10} \, M_\odot \, yr^{-1}/L_\odot$, yielding
for \sourcename~ a SFR in between 3 and $8 \times
10^3$~\Msun~yr$^{-1}$.  The massive reservoir of molecular gas of $6
\times 10^{10} \, M_\odot$ can therefore fuel star formation at the
rate implied by the far-infrared luminosity for about 10 million
years, a typical duration for a starburst in local galaxies.  The
dense warm molecular gas is thus rapidly consumed unless its mass only
represents a fraction of the total gas available in the host galaxy of
\sourcename~, perhaps in a halo, like in the case of M~82
\citep{Walter2002}.  Infalling gas could then further sustain the
starburst activity in \sourcename~ over longer periods of time.

Multiple structures such as double source, arc or ring, which are
expected for gravitational lensing, are not detected in the maps of
the non-thermal radio continuum on any scale from 10's of mas to a few
arcsec. From the radio maps, there is therefore no evidence for
lensing in \sourcename.  No correction for magnification has therefore
been applied in this paper. Deep optical and/or near-infrared
observations would be useful to further search for evidence of lensing
in \sourcename.

The VLBA observations, which resolve out structures larger than
$0\farcs15$, indicate that the 1.4~GHz radio continuum emission should
be distributed on scales larger than  1~kpc. On the
other hand, as shown by the CO observations, the molecular gas is
distributed in a region smaller than 5~kpc. If the radio
continuum and CO emission are co-spatial, the present limits could be
compatible with a scenario wherein the starburst activity in
\sourcename~ is confined within a molecular torus or disk of size
between 1 and 5~kpc.  Similar structures with sizes of typically a few
kpc are found in the cases of PSS~2322+1944 \citep{Carilli2003} and
the Cloverleaf \citep{Venturini2003}.

Using the limits to the size derived from the CO and radio
observations and assuming that the molecular gas is distributed in a
disk, the dynamical mass is $(3 < M_{\rm dym} < 10)\ \times 10^{10}
sin^{-2} i$~\Msun, where $i$ is the inclination to the line of sight.
This range is comparable to the dynamical masses derived for other
high-$z$ quasars and galaxies and to that of local $m^*$-galaxies
\citep{Genzel2003}.  The ratio of molecular to dynamical mass is
therefore estimated to be $( 2.4 > M_{\rm H_2}/M_{\rm dyn} > 0.6) \ 
{\sin^{2} i}$, indicating that a large fraction of the total mass is
in the form of molecular gas.

\section{Conclusions}
\label{sec:conclusions}

The millimetre and radio observations described in this paper provide
strong evidence that the far-infrared luminosity of \sourcename~ is
related to massive star formation.  First, the huge reservoir of dense
and warm molecular gas ($ 6 \times 10^{10} \, M_\odot$) can sustain
the star formation rate of a few thousand solar masses per year
implied by the far-infrared luminosity for about 10 million years.
Second, the radio continuum emission is found to be consistent with
the radio/far-IR correlation of star forming galaxies. The limits on
the size of the CO and radio continuum emitting region suggest that in
\sourcename~ star formation is taking place in an extended region,
which could be a torus or disk-like structure not larger than 5~kpc.
Observations of the CO emission at higher spatial resolution would be
useful to further constrain the size of the star-forming region in
\sourcename~ and thereby its dynamical mass.
  
The submillimeter and radio properties of the $z=2.58$ quasar
\sourcename, which are similar to those seen in other high-$z$
optically luminous quasars, are consistent with a coeval starburst and
AGN.  However, the relative importance of the warm dust heated by the
star formation and the dust directly heated by the AGN is still poorly
constrained.  Observations at higher frequencies are needed to analyze
the rest-frame near- to mid-infrared spectral energy distribution of
distant quasars like \sourcename.  Such observations will allow us to
derive the total infrared luminosity of high-$z$ quasars, to estimate
the contribution of the hot dust heated by the AGN, to better
constrain the star formation history in their associated host galaxies
and, thereby, to investigate further the coeval growth of stars and
massive black holes and its evolution over cosmic time.

\begin{acknowledgements}
  We thank the IRAM Plateau de Bure staff for their support, V.T.
  Junkkarinen for sending us information on the optical Keck spectrum
  of \sourcename~ before publication and A. Weiss for helpful
  comments. We also thank an anonymous referee for comments which
  improved the content of this paper.
\end{acknowledgements}

\bibliographystyle{aa} 
\bibliography{paper,database}     

\end{document}